\documentstyle{mn}
\title[Spectroscopy of the candidate luminous white dwarf in NGC1818]
{Optical spectroscopy of the candidate luminous white dwarf  
in the young LMC cluster NGC1818}

\author[M.\,R. Burleigh et~al.]
{M.\,R. Burleigh$^1$, R.\,A. Saffer$^2$, G.\,F. Gilmore$^3$ and R.
Napiwotzki$^4$\\
$^1$ Department of Physics and Astronomy, University 
of Leicester, University Rd., Leicester, LE1 7RH \\
$^2$ Department of Astronomy and
Astrophysics, Villanova University, Villanova, PA 19085, USA \\
$^3$ Institute of Astronomy, Madingley Road, Cambridge, CB3 0HA \\ 
$^4$ Dr. Remeis-Sternwarte, Sternwartstr. 7, D-96049 Bamberg, Germany \\
}
\date{July 28th 1999}
\def\TD-1{\it TD-1\rm }
\def\tkev{\thinspace{ke\kern-.15em V}}
\def\tev{\thinspace{e\kern-.15em V}}

\def\la{\mathrel{\hbox{\rlap{\hbox{\lower4pt\hbox{$\sim$}}}{\raise2pt\hbox{$<$}}
}}}
\def\ga{\mathrel{\hbox{\rlap{\hbox{\lower4pt\hbox{$\sim$}}}{\raise2pt\hbox{$>$}}
}}}
\def\Msun{\hbox{$\rm\thinspace M_{\odot}$}}
\begin{document}
\maketitle

\begin{abstract}

An optical spectrum of the Elson et~al. (1998) candidate luminous white 
dwarf in the young LMC cluster NGC1818 shows conclusively that it is
not a degenerate star. A model atmosphere 
fit gives T$_{eff}$$\approx$31,500K and log
g$\approx$4.4, typical of a garden-variety main sequence B star. However,
if it is a true LMC member then 
the star is under-luminous by almost three magnitudes. 
Its position in the cluster colour-magnitude 
diagram also rules out the 
possibility that this is an ordinary B star.  
The luminosity is, however,  
consistent with a $\sim$0.5$\Msun$ post-AGB
or post-EHB object, although if it has evolved via single star evolution
from a high mass (7.6$-$9.0$\Msun$) progenitor then 
we might expect it to have a much higher mass, $\sim$0.9$\Msun$. 
Alternatively, it has evolved in a close binary.
In this case the object offers no
implications for the maximum mass for white dwarf progenitors, or the
initial-final mass relation. Finally, we suggest that it could in fact be
an evolved member of the LMC disk, and merely projected by chance onto NGC1818.
Spectroscopically, though, we cannot distinguish 
between these evolutionary states without higher resolution (echelle) data. 

\end{abstract}


\section{Introduction}

White dwarfs are believed to be the final endpoint of stellar evolution
for all stars M$\la$8$\Msun$ (Weidemann 1987). Above some critical
mass, M$_c$, single 
stars end their lives by detonating as Type II supernovae, leaving
neutron star or black hole 
remnants. However, there are few observational constraints
on M$_c$. For example, the earliest type star known with a white dwarf
companion is the B5V star y Pup (HR2875, Vennes, Bergh\"ofer and
Christian 1997, Burleigh \& Barstow 1998). 
The degenerate star in this binary system  
must have evolved from a progenitor
with a mass greater than that of its main sequence companion, 
6.0$-$6.5$\Msun$. 

A lower limit on this maximum mass for white dwarf
progenitors can also be derived from the study of white dwarfs in young
galactic open clusters. 
Observations of four white dwarfs in the Milky Way cluster
NGC2516 by Koester \& Reimers (1996) 
imply the initial stellar mass for forming a white dwarf there is
$\approx$7$\Msun$, although Jeffries (1997) used the cluster metallicity to
revise and decrease these particular progenitor masses to only
5$-$6$\Msun$. 

Recently, 
Elson et~al. (1998) announced the discovery of a candidate 
luminous white dwarf 
in the young cluster NGC1818 in the Large Magellanic Cloud (LMC). 
This cluster has a main sequence turn-off mass of between 7.6$\Msun$ and 
9.0$\Msun$,
depending on whether convective core overshoot is assumed in the models.
If this object is indeed a young, hot, massive white dwarf, then the
lower limit for the maximum mass for white dwarf progenitors (M$_c$)
would be $\ga$7.6$\Msun$. 

Elson et~al. (1998) used U, V and I colours (obtained with
WFPC2 on the Hubble Space Telescope) to identify their white dwarf
candidate. However, as pointed out by Liebert (1999), at V$\approx$18.4
this object is highly unlikely to be a young, hot white dwarf. 
At the
distance of the LMC, $\sim$50kpc, the star has an absolute visual
magnitude near zero. The most luminous white dwarfs in the Palomar Green
Survey, though,  
are fainter than M$_v$$\sim$6, and in NGC1818 would be no brighter
than V$\sim$24.5 (Green, Schmidt and Liebert 1986). 
In addition, 
clusters with main sequence masses $\ga$5.0$\Msun$ 
produce massive white
dwarfs, $\ga$0.9$\Msun$. By implication these degenerate stars have
abnormally small radii, and at the distance of the LMC would appear at
V$\ga$29. 

\begin{figure*}
\vspace{8.5cm}
\includegraphics{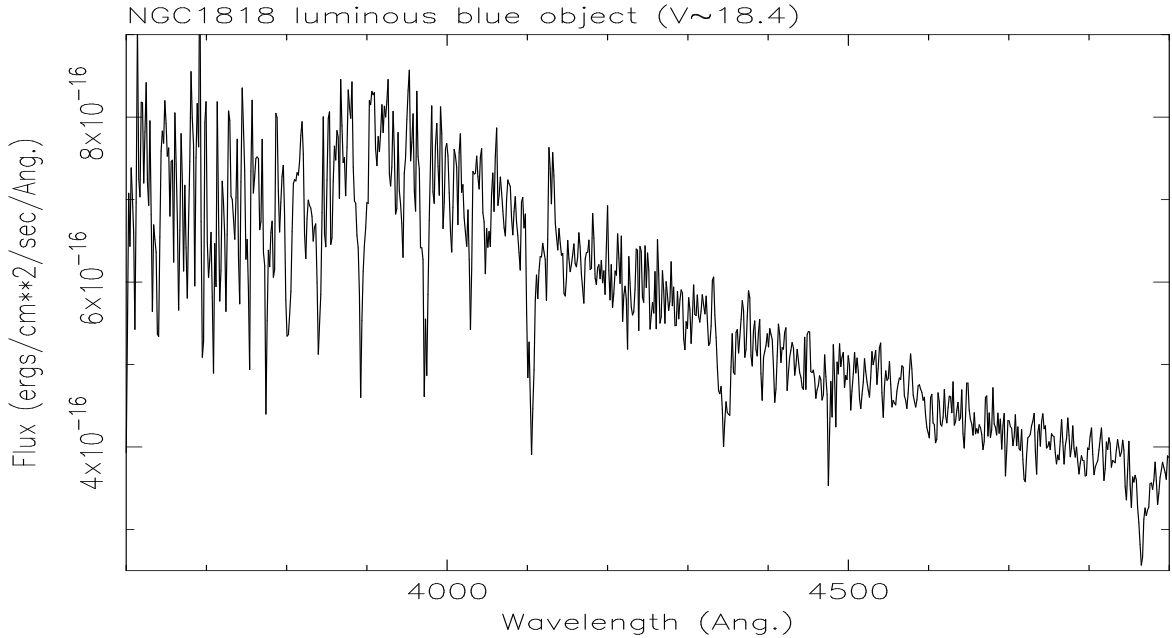}
\caption{Optical spectrum of the luminous blue object in NGC1818}
\end{figure*}

\begin{figure}
\vspace*{11cm}
\includegraphics{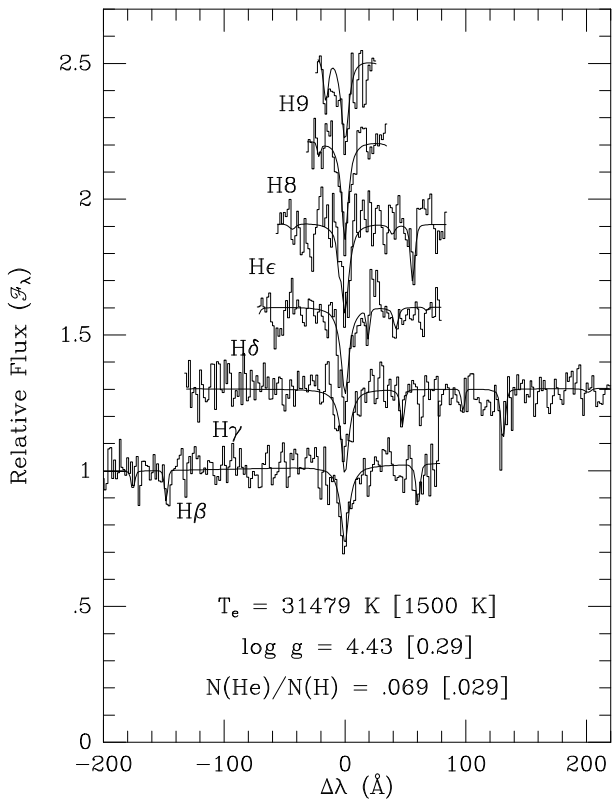}
\caption{Model fit to the H Balmer lines}
\end{figure}

In this paper we present an optical spectrum of the Elson et~al. 
object, and show conclusively that it is not a white dwarf. 
We argue that, although spectroscopically it resembles a 
garden-variety main sequence B star, 
it appears to be under-luminous by almost three magnitudes. 
Its position in the cluster HR diagram also excludes this possibility.  
Instead, it may be an object 
evolving off the extended horizontal branch or, less likely, a post-AGB
star. It may even have evolved in a close binary system. However, without 
high resolution echelle spectroscopy, it is impossible to distinguish
between these evolutionary states. 

\section{Optical observation}

The white dwarf candidate was observed for 1800 secs 
with the 3.9m Anglo-Australian
Telescope (AAT) on 1998 March 5th, using the RGO spectrograph and 600V
grating centred at 4200{\AA}. This gave a dispersion of
$\approx$1.55{\AA} per pixel on a Tek CCD (running at 170K),
equivalent to a resolution of $\approx$4{\AA} (FWHM). 
A blue/UV flux calibration standard was also observed. 
These data (Figure 1) 
were reduced with the Starlink package FIGARO. 
We note that sky subtraction was difficult, due to the high density of
stars in this region, and that the seeing was poor, $>$1.5 arcseconds. 

A nearby comparison star was also observed. The location of this object
on the main sequence in the cluster 
colour-magnitude diagram makes it a
highly probable cluster member (see Figure 1 of Elson et al. 1998). 

\begin{figure*}
\vspace*{9cm}
\includegraphics{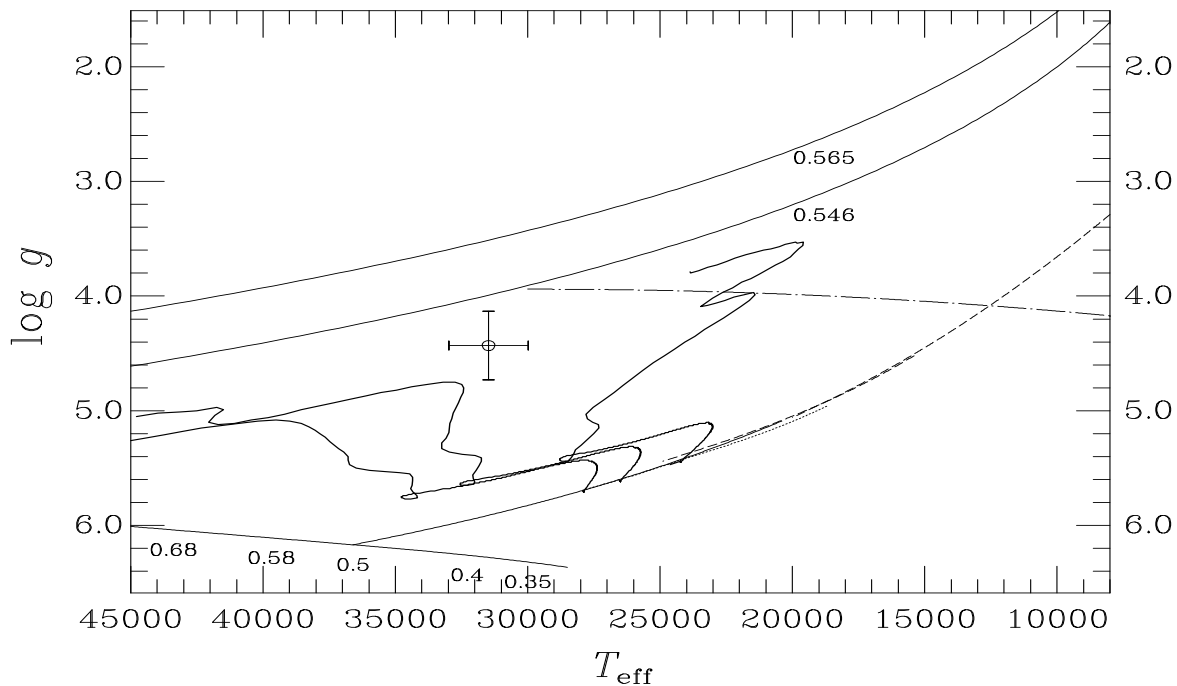}
\caption{The position of the NGC1818 object in the T$_{eff}$$-$log g
plane. Also plotted are evolutionary
tracks from Sch\"onberner (1993) for post-AGB stars (solid diagonal lines;
core masses labeled in units of 1$\Msun$), the zero-age EHB 
from Sweigart (1987, dashed line), and the zero-age main sequence
(dot-dashed line).
Loci showing how objects with a variety of masses
evolve away from the EHB are also shown (Caloi 1989).}
\end{figure*}

\section{Analysis}

The spectrum is shown in Figure 1. Although the continuum rises fairly
steeply towards the blue, indicating that the object is hot, it is
immediately obvious that the H Balmer lines are far too narrow
for this to be a high gravity object, such as a white dwarf. HeI absorption 
lines are also visible,
for example at 4471{\AA} and 4026{\AA}, along with CaII at 3933.7{\AA}. 

A model atmosphere fit to the line profiles (Figure 2) gives
T$_{eff}$ $=$31,500K$\pm$1500K, log g$=$4.4$\pm$0.3, and
He/H$=$0.07$\pm$0.03. These parameters are consistent with a 
garden-variety main sequence 
B star, albeit with a slightly low He abundance which is
not very significant.  

We note that the difference in radial velocity between this object and
the comparison star, measured by cross-correlation, is 10$\pm$15 km
sec$^{-1}$. Therefore, it is almost certainly a member of the LMC.
However, the expected internal velocity dispersion in the NGC1818 cluster
is only $\sim$1 km sec$^{-1}$. Thus we would only be able to test for cluster
membership, as opposed to LMC membership, with high dispersion (echelle) 
spectra.

\section{Discussion}

The optical spectrum of the luminous blue object in the LMC cluster 
NGC1818 shows conclusively
that it is not a white dwarf. Our model fit shows that it could be a
normal main sequence B star, T$_{eff}$$\approx$31,500K, log
g$\approx$4.4.  

However, the position of this star in the
T$_{eff}$$-$log g plane is ambiguous since, in addition to the main
sequence, 
post-asymptotic giant branch (post-AGB) and post-extended horizontal branch
(post-EHB) tracks all cross the same area. 

Figure 3 shows the T$_{eff}$$-$log g plane and the position of the
NGC1818 object (marked by the large cross). Also plotted are evolutionary 
tracks from Sch\"onberner (1993) for post-AGB stars (solid diagonal lines;
core masses labelled in units of 1$\Msun$), the zero-age EHB 
from Sweigart (1987, dashed line), and the zero-age main sequence
(dot-dashed line). 
Loci showing how objects with a variety of masses
evolve away from the EHB are also shown (Caloi 1989). Obviously, 
there are a number of plausible alternative interpretations for the
nature of this object. We now discuss each possibility in turn:

\vspace{3mm}
\noindent
A) This is a normal main sequence B star, lying at the 
distance of NGC1818 and probably a genuine cluster member. Rapid rotation
might be expected, with atmospheric abundances characteristic of the
cluster. At V$\approx$18.4, though, it would be impossible to determine these
parameters without an 8- or 10-m class telescope. 

We note, however, that this star appears to be 
under-luminous by almost three magnitudes  
for a B star at the distance of the LMC. A 31,500$\pm$1500K zero-age main 
sequence star 
in the LMC (Z$=$0.008) has a gravity log g$\leq$4.33 and an absolute
magnitude of $-$2.8$\pm$$-$0.4, (Schaerer et~al. 1993), 
yet our V$=$18.4 star has an absolute
magnitude of only $\approx$0 (assuming a distance of 50kpc).  
In addition, the position of this object
bluewards of the main sequence in the cluster colour-magnitude diagram (see
Figure 1 of Elson et~al., 1998) excludes the possibility that this is an
ordinary B star. 
Therefore, we must seriously consider other evolutionary states for this
hot object.

\vspace{3mm}
\noindent
B) This is an object on its way to
becoming a white dwarf, i.e. a post-AGB star. 
Figure 3 demonstrates that low-mass (e.g.~0.546$\Msun$)  
post-AGB tracks run through the same area of the HR
diagram as a 31,500K, log g$=$4.4 B star. 
%However, although the luminosity of this object 
%is consistent with a low mass
%post-AGB star, it could be argued that a mass of $\approx$0.55$\Msun$ 
%is too low for a
%star to ever reach the AGB, at least via single star evolution alone.
However, Liebert (1999) argues that this star is unlikely to be a
post-AGB cluster member. As pointed out in the introduction to
this paper, we would expect its mass to be high ($\ga$0.9$\Msun$)
compared to typical, older stellar remnants. A high mass post-AGB star,
though, would have a luminosity inconsistent with the NGC1818 object. For
example, a 0.855$\Msun$ post-AGB star  
has a luminosity log L$/$L$_\odot$$\sim$4.3 (Vassiliadis \& Wood 1994),
yet our object has a luminosity of only log L$/$L$_\odot$$\sim$3.0. 

\vspace{3mm}
\noindent
C) This is a post-EHB cluster member. The timescale for evolution
in this phase certainly 
makes it more likely that this is a post-EHB object rather
than a post-AGB star. Sch\"onberner (1983) gives the timescale for
post-AGB cooling in this region as 10$^3$$-$10$^4$ years, while that for
HB evolution through this region is 10$^6$$-$10$^7$ years (Castellani
et~al. 1994). The surface gravity is as expected for a
post-EHB object, although it is too low for an object on the zero-age EHB 
(at this temperature, see Figure 3). The luminosity is also consistent
with a $\sim$0.5$\Msun$ post-EHB object. 
However, as with the post-AGB scenario, the formation of
such an object via single star evolution may be unlikely, since the star
would have had to lose around 6$\Msun$ of envelope as it ignited helium.
Horizontal branch stars, like post-AGB stars, are slow rotators, 
so high resolution spectra would help to distinguish its evolutionary
status. If it is a post-EHB star, then during its
time at high gravity prior to He-exhaustion it would have
altered its abundances through diffusion. Again, though, there is no way
to tell without much higher resolution data. 

\vspace{3mm}
\noindent
D) Liebert (1999) offers one other speculative interpretation. Perhaps this
object has been formed through close binary evolution, such that the
white dwarf progenitor has lost its envelope (through mass transfer to the
companion) before the mass of the core has reached the level required for
helium ignition. 
This undermassive progenitor core would evolve on a post-RGB 
track that is parallel to, but at a much lower luminosity than, the
post-AGB track for any higher mass core produced by single star
evolution. Such systems have been observed, for example, in the centres
of planetary nebulae (Napiwotzki 1999). 
In this case the object is indeed becoming a white dwarf, but
because it has been evolving via binary evolution it 
offers no implications for the upper mass limit for white dwarf
progenitors. 

%After traversing the HR diagram to the left, it would enter
%the white dwarf cooling sequence as one of the $\sim$10\% of stars in the
%low mass tail of the Bergeron, Saffer \& Liebert (1992) mass
%distribution. As a member of a cluster with a high turn-off mass, its
%companion might be a {\em very} massive white dwarf or neutron star. If
%the binary period is short enough, this pair might eventually become
%e.g.~an AM CVn binary, low mass x-ray binary, or even a Type Ia or Ib
%supernova.

However, if it is in a close binary then it should be
suffering large radial velocity variations and, statistically, we would
expect to see it near velocity extrema. Since the velocity appears to be
the same as the LMC velocity, we might conclude that close binary evolution
is a low-probability alternative. Again, though, our current 
data may be too low in resolution to draw such a conclusion. 

\vspace{3mm}
\noindent
E) Finally, we suggest that this object could be a post-EHB star, but
that it is not a member of the NGC1818 cluster. Instead, it lies 
in the disk of the LMC and simply appears projected by chance onto
NGC1818. Although this may seem statistically unlikely, we note that
considerable star formation has occured in the LMC over the last $\sim$2
Gyr and thus the object need not have evolved from a high mass
progenitor. It also need not have formed
through binary evolution. However, 
no convincing horizontal branch has been observed in the vicinity of
NGC1818 (see Fig. 4 of Hunter et~al. 1997), and thus the existence of a 
post-EHB star in this region is improbable.    
Once again, though, the test of this option is an
echelle spectrum with much improved velocity resolution. 

\vspace{3mm}

A comparison can be drawn between this object and the V$\approx$14.5 hot 
blue star 
PG~0832$+$676 (Hambly et~al. 1996). In low resolution spectra this
object also closely resembles a young B-type star. However, high
resolution observations demonstrate that it is extremely sharp lined ($v$
sin $i \sim$ 1 km sec.$^{-1}$), has a low projected rotational velocity, 
and that although the abundances of
helium, nitrogen and oxygen are near normal, there is a systematic
depletion of $\sim$0.4 dex in the abundances of other elements. 
Therefore, Hambly et~al.  concluded that the object is
most likely an old, evolved star, either in the post-AGB phase or more
probably evolving off the EHB. 

High resolution spectroscopy is similarly now required for the NGC1818 object, 
for a detailed abundance analysis, a rotational velocity
determination, and an accurate radial velocity determination, in order to
distinguish between the various possible evolutionary states. However, at
V$=$18.4 this will only be possible with the new generation of 
southern hemisphere 
8$-$10-m class telescopes, such as the VLT or Gemini.

\section*{Acknowledgements}

MRB acknowledges the support of PPARC, UK. We thank Sabine Moehler of the 
Dr. Remeis Sternwarte, Bamberg, Germany, for useful comments on the nature of 
this object. We thank Helen
Johnston and Brian Boyle of the Anglo-Australian Observatory for their
help in obtaining the spectrum presented here.

\end{document}